\documentclass{PoS}

\title{The latest T2K neutrino oscillation results}

\ShortTitle{latest T2K neutrino oscillation results}

\author{\speaker{Le\"ila Haegel}\thanks{For the T2K collaboration.}\\
        University of Geneva, Switzerland\\
        E-mail: \email{leila.haegel@cern.ch}}

\abstract{T2K is a long-baseline neutrino oscillation experiment taking data since 2010. A neutrino beam is produced at the J-PARC accelerator in Japan and is sampled at a Near Detector complex 280 m from the neutrino production point and at the far detector, Super-Kamiokande, located 295 km from the source. Beams predominantly composed of muon neutrinos or muon anti-neutrinos have been produced by changing the currents in the magnetic focusing horns. This presentation will show the most recent T2K oscillation results obtained from a combined analysis of the entire available data set in the muon neutrino and muon anti-neutrino disappearance channels, and in the electron neutrino and electron anti-neutrino appearance channels. The data cover runs 1 to 8 (2010 to 2017) and consist of $7.252 \cdot 10^{20}$ POT in neutrino mode and $7.531 \cdot 10^{20}$ POT in antineutrino mode. Using these data, we measure four oscillations parameters: $\sin^2 \theta_{23}$, $\sin^2 \theta_{13}$, $\Delta m_{32}^2$ and $\delta_{CP}$. The analysis excludes CP-conservation in the neutrino sector at 90\% C.L.}

\FullConference{The European Physical Society Conference on High Energy Physics\\
		5-12 July, 2017\\
		Venice}

\begin{document}

\section{The T2K experiment}

\begin{figure}
  \centering
    \includegraphics[width=0.4\textwidth]{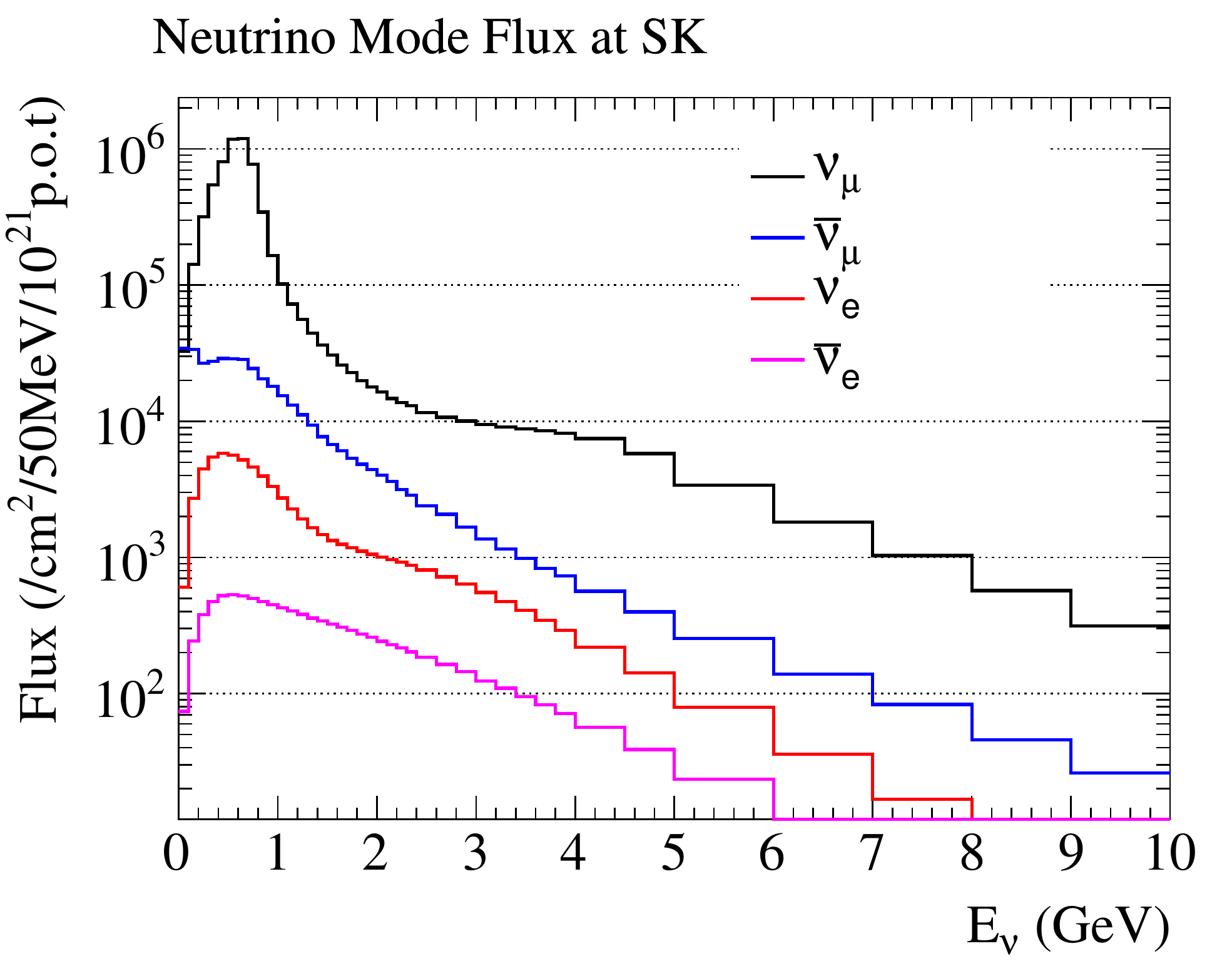}
    \includegraphics[width=0.4\textwidth]{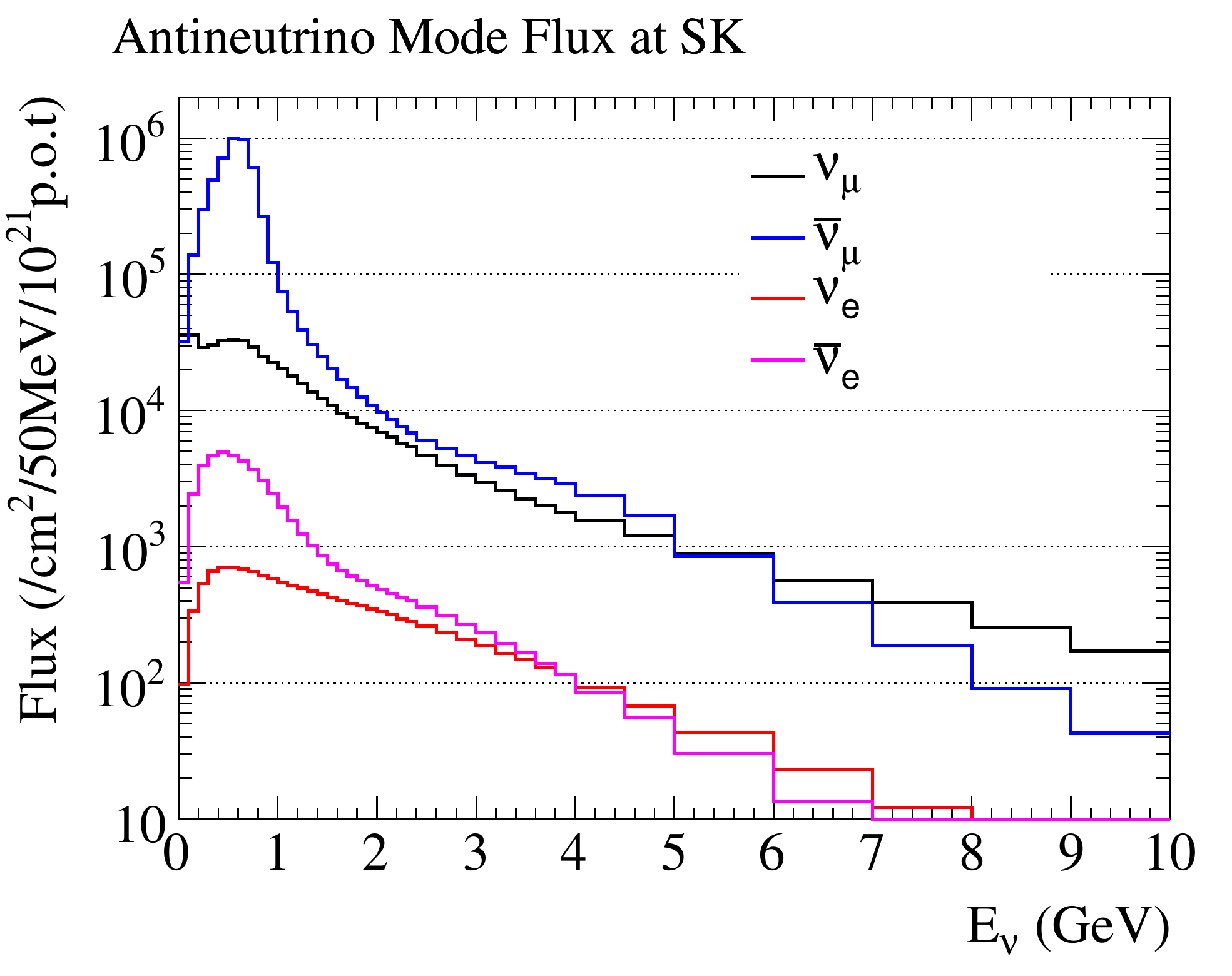}
  \caption{Expected flux in the far detector Super-Kamiokande in neutrino mode (left) and antineutrino mode (right). The stacked histograms represent the particle content. }
  \label{t2k_flux}
\end{figure}

The T2K experiment is a long-baseline neutrino oscillation experiment located in Japan~\cite{t2k_exp}. 
A neutrino beam is created in J-PARC from a 30 GeV proton beam hitting a carbon target, creating charged pions and kaons. 
Three horns select particles according to their electric charge, so a positive horn current selects positively charged particles decaying into $\mu^+ + \nu_{\mu}$ in order to deliver a neutrino-enhanced beam (refered to as neutrino mode) ; and a negative horn current selects negatively charged particles decaying into $\mu^- + \bar{\nu_{\mu}}$ to deliver an antineutrino-enhanced beam (refered to as antineutrino mode). 
A negligible amount of mesons decay into $\nu_{e}$ and $\bar{\nu_{e}}$, providing a contamination less than 1\%.
The neutrino beam energy peaks at 600 MeV as shown on Figure~\ref{t2k_flux}, and its direction is monitored on a daily basis by the on-axis INGRID detector located 280m downstream.

\begin{figure}
  \centering
    \includegraphics[width=0.4\textwidth]{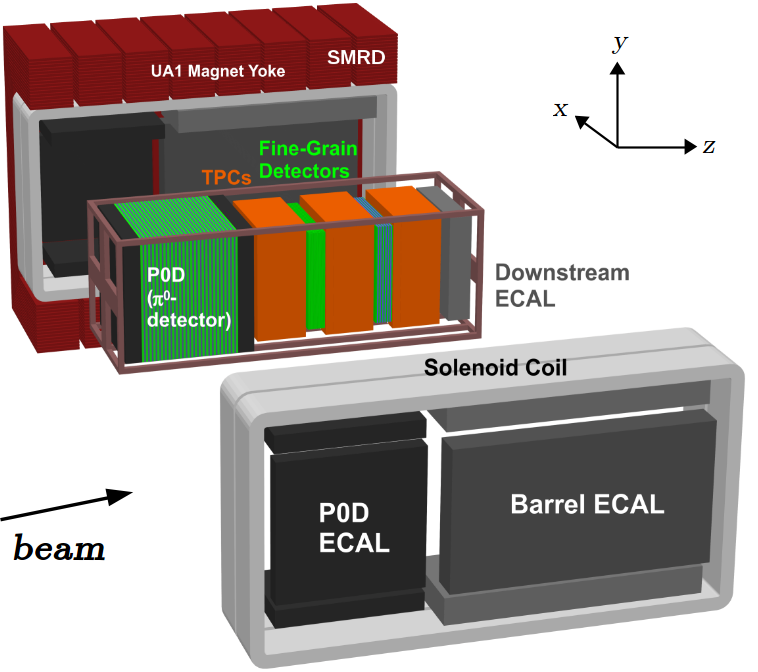}
    ~
    \includegraphics[width=0.55\textwidth]{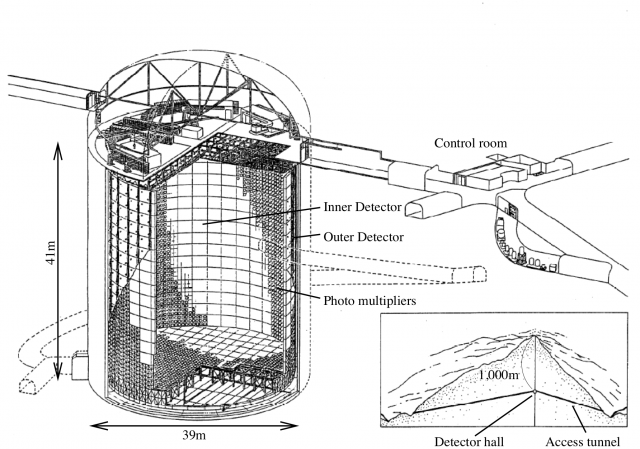}
  \caption{The off-axis near (left) and far (right) detectors of the T2K experiments. }
  \label{t2k_det}
\end{figure}

Above INGRID is located the off-axis near detector ND280, made of several subdetectors embedded in the UA1 magnet providing a 0.2 T horizontal magnetic field as shown on Figure~\ref{t2k_det} left. 
The Fine Grained Detectors (FGDs) are the two target detector, the first being composed of fifteen modules, each module containing two layers of extruded polystyrene scintillator bars placed consecutively along $\vec{x}$ then $\vec{y}$
to ensure the tracking of the charged particles. 
The second FGD contains eight similar modules, interlaid with water modules in order to select interactions on oxygen. 
The two FGDs are placed in between three Time Projection Chambers (TPCs) filled with a gas mixture mainly composed of argon and holding a central cathode creating a linear electric field in the $\vec{x}$ direction. 
The charged particles entering the TPC ionise the gas, and can be identified from their energy loss combined with the momentum measurement from the curvature of the particle track in the magnetic field.
The $\pi^0$ detector (P$\varnothing$D), electromagnetic calorimeters (ECals) and Side Muon Ranger Detector (SMRD) aim at constraining the background interactions and veto the particles originating from outside the FGDs.

The far detector Super-Kamiokande (Super-K) is located at the first (anti-)$\nu_e$ appearance maximum, 295 km from J-PARC and 2.5$^{\circ}$ off-axis from the beam center in order to reduce the energy width of the neutrino flux. 
It consists of a 50 kton cylindrical detector pictured on Figure~\ref{t2k_det} right, separated in an outer and inner detector both filled with water. 
Charged-current (CC) interactions of neutrinos with oxygen nucleons create charged leptons, that cross the detector while creating a Cherenkov cone of light detected by the 11,146 photomultipliers covering 40\% of the inner detector wall. 
The fuziness of the Cherenkov rings enables  $\mu^{+/-}$ to be separated from $e^{+/-}$, giving an indication on the flavour of the neutrino having produced the interaction (but without providing information on the electric charge of particles).

\section{The parameter estimation methodology}

The analysis aims at measuring four of the six parameters of the PMNS matrix, namely $\sin^2 \theta_{23}$, $\sin^2 \theta_{13}$, $\Delta m_{32}^2$ and $\delta_{CP}$, by combining (anti-)$\nu_{\mu}$ disappearance and (anti-)$\nu_{e}$ appearance data in neutrino and antineutrino modes~\cite{t2k_oa5}.
The oscillation parameters are estimated from samples of oscillated neutrino interactions selected in Super-Kamiokande.
Three samples of CC interactions are selected in neutrino mode: one sample of $\nu_{\mu}$ and one sample of $\nu_{e}$ charged-current quasi-elastic (CCQE) interactions, that only produce a outgoing nucleon under the Cherenkov threshold, and a charged lepton whose kinematic informations are sufficient to extract the energy of the incident neutrino with the equation: 
\begin{equation}
\label{eq:SK_Erec}
E_{\nu}^{CCQE} = \frac{m^2_f - (m_i')^2 - m^2_l + 2m_i' E_l}{2(m_i' - E_l + p_l\cos\theta_l)}
\end{equation}
\noindent where $E_{\nu}^{CCQE}$ is the reconstructed neutrino energy assuming a CCQE interaction, $m_i$ and $m_f$ are the initial and final nucleon masses respectively, $m_i' = m_i - E_B$, where $E_B=27$ MeV is the binding energy of a nucleon inside $^{16}\mathrm{O}$ nuclei, and $E_l$, $p_l$ and $\theta_l$ are the lepton energy, momentum and angle with respect to the incident neutrino angle respectively. 
The third sample contains $\nu_{e}$ charged-current charged-pion production (CC-1$\pi^+$), where an electron is produced with an undetected nucleon and a charged pion, that is selected to be under the Cherenkov threshold and tagged from delayed clusters of photomultipliers hits. 
Assuming that the reaction occurs with the production of a $\Delta^{++}$ resonance decaying into a pion and nucleon, the neutrino energy can be infered with the equation: 
\begin{equation}
\label{eq:SK_CC1pi_Erec}
E_{\nu}^{\Delta} = \frac{m^2_{\Delta^{++}} - m^2_p - m^2_l + 2m_pE_l}{2(m_p - E_l + p_l\cos\theta_l)}
\end{equation}
\noindent where $E_{\nu}^{\Delta}$ is the reconstructed neutrino energy assuming a $\Delta^{++}$ production interaction, and $m_{\Delta^{++}}$ is the mass of the $\Delta^{++}$. 
In antineutrino modes, two CCQE samples of $\bar{\nu_{\mu}}$ and $\bar{\nu_{e}}$ interactions are selected. 
The analysed data cover runs 1 to 8 (from 2010 to 2017) and consist of $7.252 \cdot 10^{20}$ POT in neutrino mode and $7.531 \cdot 10^{20}$ POT in antineutrino mode. 
Using the prior values of the oscillation parameters given on Table~\ref{tab:ch04:evts_skmcdata} left, the predicted number of Super-K events is given on Table~\ref{tab:ch04:evts_skmcdata} right. 
The right side of Table~\ref{tab:ch04:evts_skmcdata} also shows the observed number of Super-K events, exhibiting an excess of observed $\nu_{e}$ and a deficit of observed $\bar{\nu_{e}}$.
\begin{table}[ht!]
\small
\centering
\parbox{.3\linewidth}{  
\centering
\begin{tabular}{ c c}
  \hline \hline
          Parameter & Value \\
          \hline
           $\sin^2 2 \theta_{12}$ 	& 0.846   \\
          $\Delta m^2_{21}$   	& $7.53 \times 10^{-5}$ eV$^2 \cdot$c$^{-4}$   \\
          $\sin^{2}\theta_{23}$ & 0.528 \\
           $\Delta m^2_{32}$ & $2.509\times 10^{-3}$ eV$^2 \cdot$c$^{-4}$   \\
           $\sin^{2}2\theta_{13}$ & 0.085 \\
           $\delta_{CP}$ & -1.601 \\
\hline \hline
        \end{tabular}
}
\hspace*{.5cm}
\parbox{.65\linewidth}{
\centering
\begin{tabular}{lccccc}    
\hline \hline
& $\nu_{\mu}$& $\nu_{e}$&  $\nu_{e}$& $\bar{\nu_{\mu}}$& $\bar{\nu_{e}}$\\
    & CCQE & CCQE & CC-1$\pi^+$ & CCQE & CCQE \\
\hline
    predicted            & 136.214        & 28.748   & 3.217      & 64.404      & 6.011       \\
    observed             & 135            & 32       & 5          & 66          & 4           \\
\hline \hline
\end{tabular}
}
\caption{Left: prior value of the oscillation parameters.
Right: predicted and observed total number of events for the five Super-K samples in the oscillated case. }
\label{tab:ch04:evts_skmcdata}
\end{table}

The systematical uncertainty on the number of events ranges from $\sim$ 12\% (the four CCQE samples) to $\sim$ 20\% (the CC-1$\pi^+$ sample where the uncertainty on the particle identification is larger).
In order to decrease the systematic uncertainty due to flux and cross-section modelling, a fit of the unoscillated ND280 data is performed.
Seven samples are selected in ND280 for this purpose: three samples of $\nu_{\mu}$ CCQE, CC-1$\pi^+$ and other CC interactions in neutrino mode, two samples of $\bar{\nu_{\mu}}$ CCQE and non-CCQE interactions in antineutrino mode, as well as two samples of $\nu_{\mu}$ CCQE and non-CCQE interactions in antineutrino mode to constrain the strong contamination of neutrino events in antineutrino mode visible on Figure~\ref{t2k_flux}. 
After including the ND280 information, the total statistical uncertainties on the numbers of events are reduced by a factor of three for the CCQE samples and are reduced by a factor of two for the CC-1$\pi^+$ sample.

Two statistical methods are used to estimate the neutrino oscillation parameters. 
The first is a frequentist method, that consists of first fitting the ND280 data, then propagating the obtained constraints on the flux and cross-section parameters in a covariance matrix used as prior on the fit to the Super-K data. 
A minimising log-likelihood method is used to estimate the value and confidence intervals of the oscillation parameters, with the Feldman-Cousins method applied on the extraction of $\delta_{CP}$~\cite{feldman_cousins}. 
The second method relies on Bayesian statistics, performing a simultaneous analysis of the ND280 and Super-K data where the posterior probability density is sampled with a Markov Chain Monte-Carlo method in order to extract credible intervals and best-fit values of the oscillation parameters. 
In both cases, the likelihood is marginalised over the systematic parameters and the analysis is performed with and without the information of $\sin^2 \theta_{13}$ measured by the reactor experiments~\cite{pdg}.

\section{Results of oscillation parameter estimation}

\begin{figure}[ht!]
  \centering
  \includegraphics[width=0.4\textwidth] {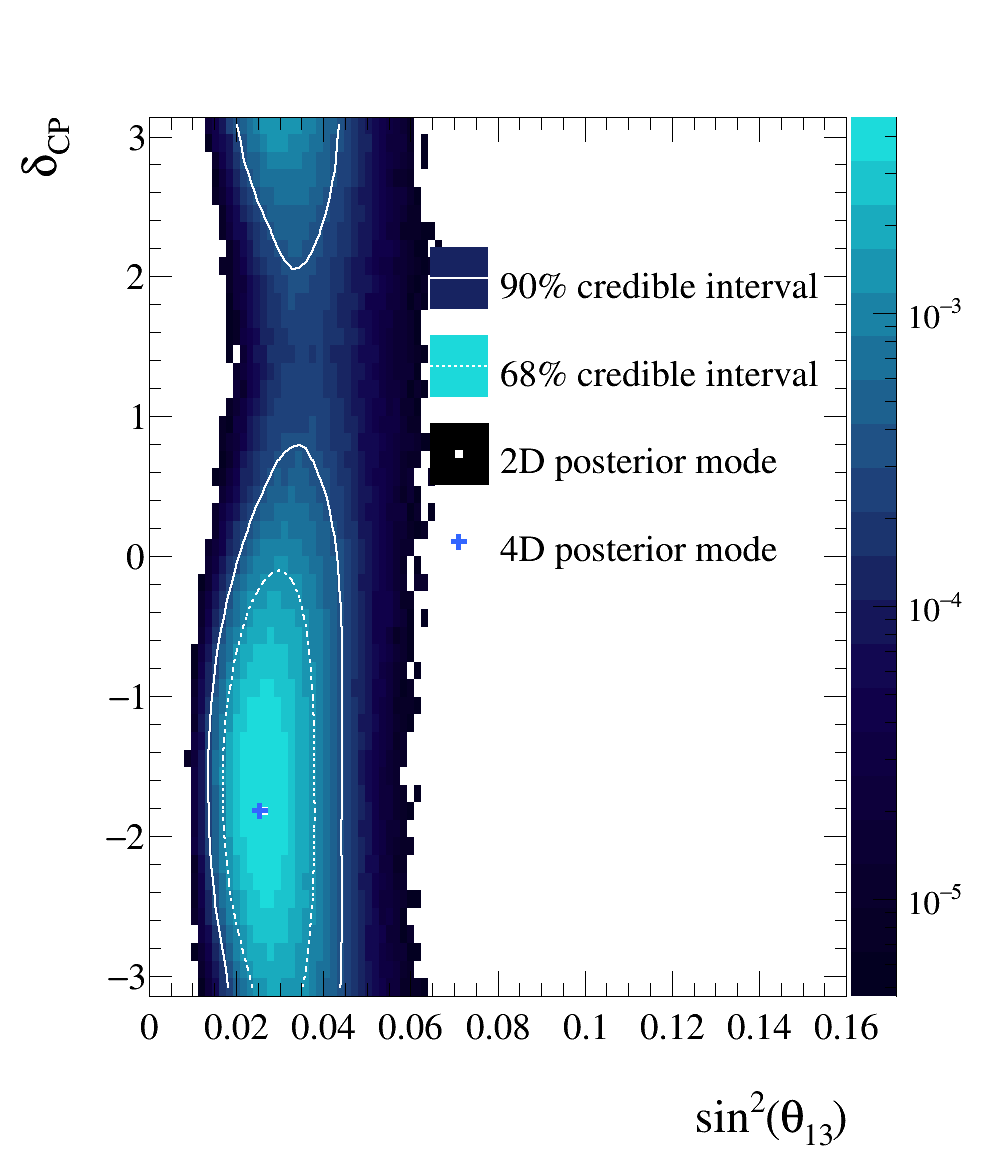}
    \includegraphics[width=0.59\textwidth]{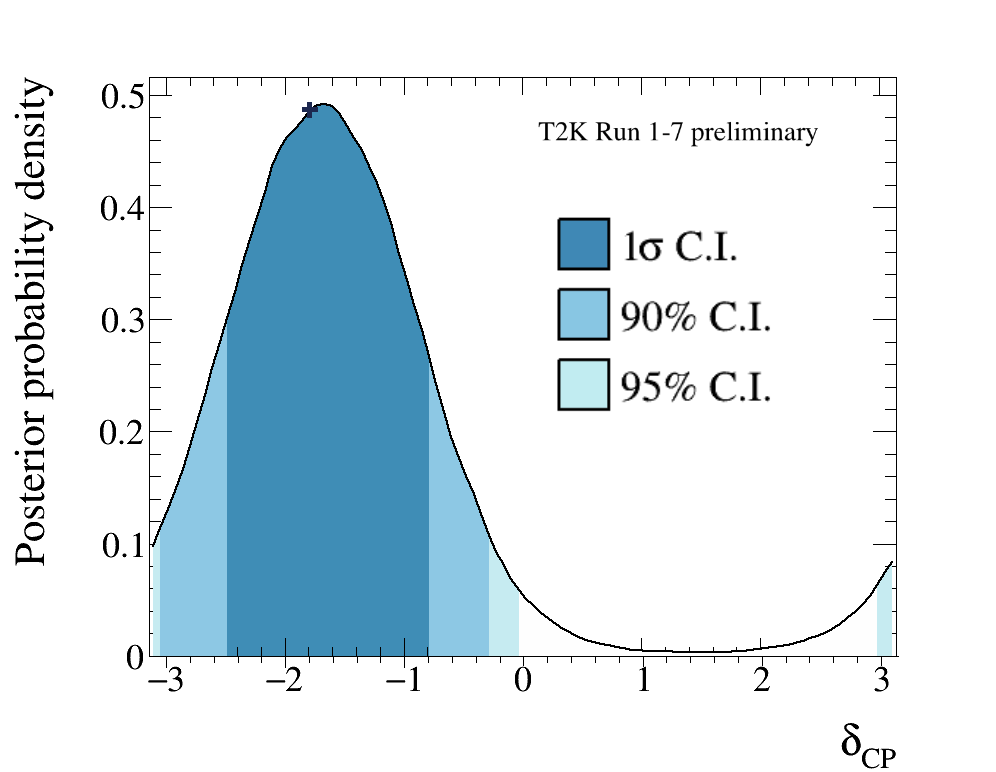}
  \caption{
  Left: posterior probability density in the $\sin^2 \theta_{13}$-$\delta_{CP}$ space, without the reactor information on $\sin^2 \theta_{13}$ from the reactor experiments.  
  Right: posterior probability density on $\delta_{CP}$, including the reactor information on $\sin^2 \theta_{13}$, where the cross represent the best-fit without marginalisation over $\sin^2 \theta_{13}$, $\sin^2 \theta_{23}$ and $\Delta m_{32}^2$.
  Both figures includes both neutrino mass orderings. }
  \label{t2k_results}
\end{figure}

When using the T2K data only, the best fit of $\sin^2 \theta_{13}$ is found higher than the reactor value as shown on Figure~\ref{t2k_results}, however the measurement is compatible with both experiments as the $\sin^2 \theta_{13}$ results from the reactor experiments is inside the 1$\sigma$ interval.
A separate analysis of the $\nu_{e}$ and $\bar{\nu_{e}}$ data has shown that the high value of $\sin^2 \theta_{13}$ is driven by the $\nu_{e}$ data, while the $\bar{\nu_{e}}$ data agree better with the reactor value.

When including the reactor information on $\sin^2 \theta_{13}$, the CP-violating phase $\delta_{CP}$ is found consistent with maximal CP violation ($\delta_{CP}=-\pi/2$) at 1$\sigma$. 
The CP-conserving values of \linebreak $\delta_{CP}=\{-\pi;0;+\pi \}$ are excluded at 90\% in the credible intervals shown on Figure~\ref{t2k_results} right, that includes both neutrino mass orderings. 

\begin{figure}[ht!]
  \centering
    \includegraphics[width=0.6\textwidth]{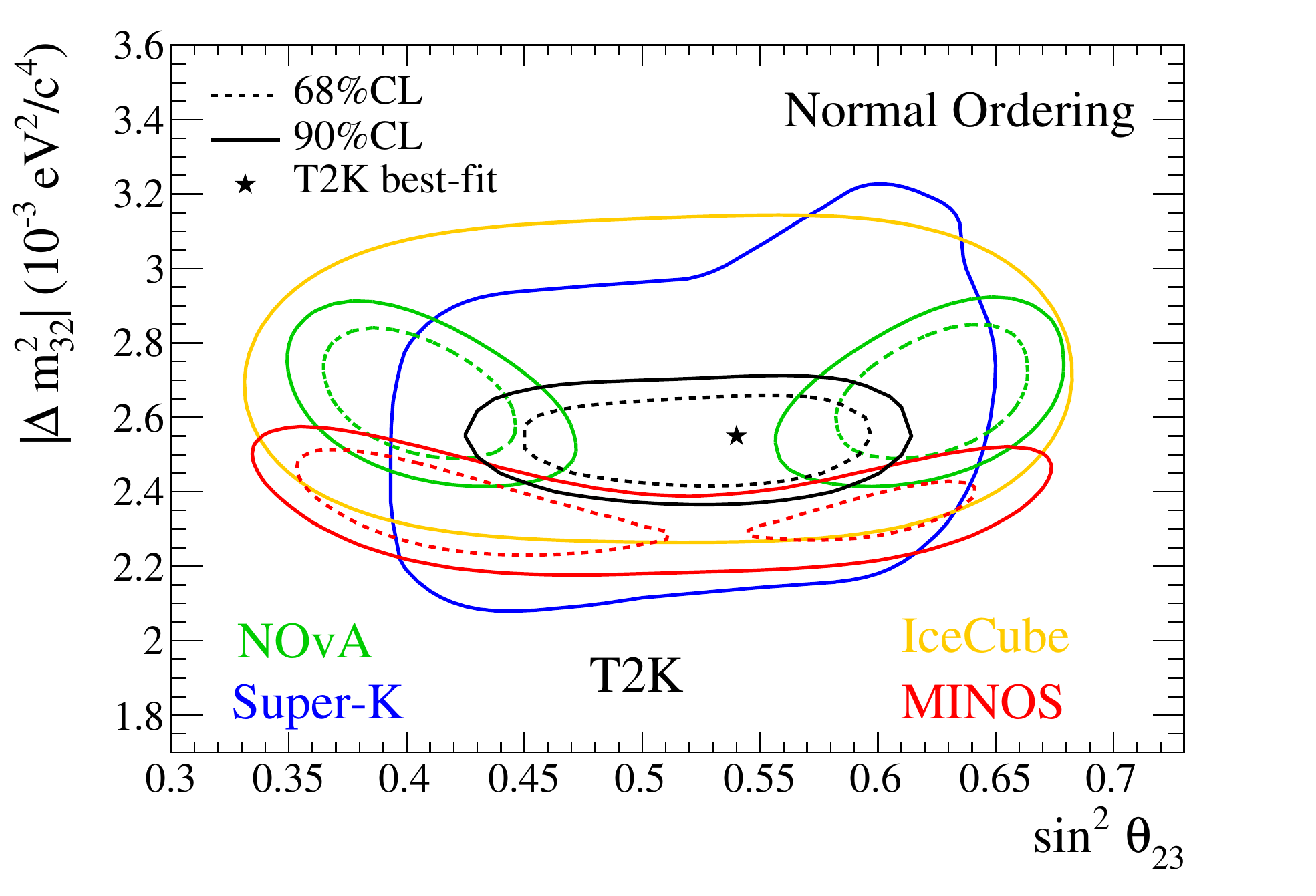}
  \caption{The 90\% and 68\% confidence levels in the $\sin^2 \theta_{23}$-$\Delta m_{32}^2$ space from T2K compared to other experiments, assuming $\Delta m_{32}^2>0$ (normal ordering of neutrino masses).}
  \label{t2k_resultsd}
\end{figure}

As shown on Figure~\ref{t2k_resultsd} left, $\sin^2 \theta_{23}$ is found consistent with maximal (anti-)$\nu_{\mu}$ disappearance, corresponding to $\sin^2 \theta_{23}=0.5$.
Assuming the normal mass hierarchy, $\Delta m_{32}^2$ is found lower than the NO$\nu$A value, that excludes maximal (anti-)$\nu_{\mu}$ disappearance at 90\% confidence level, and higher than MINOS. 
The results are however in agreement in the 90\% confidence level between all the experiments.

The probability that the neutrino mass ordering is normal, i.e. $\Delta m_{32}^2>0$, is found to be 0.788, leading to a Bayes factor of 3.72.
Similarly, the Bayes factor that $\sin^2 \theta_{23}>0.5$, also refered to as the higher octant, is found to be 2.41. 
Both of those values are only indications of a preference of the T2K data for the normal mass ordering and higher octant, as they are too low to be conclusive.

\section{Conclusion and future prospects}

T2K has excluded CP-conservation at 90\% confidence levels, and measured the neutrino oscillation parameters as summarised on Table~\ref{tab:bestfit_worc_run1-7c_mach3}.
\begin{table}[ht!]
\centering
\begin{tabular}{ l c c  c  c  }\hline \hline
 Parameter               & Best-fit                            &  $\pm1\sigma$     \\ \hline 
 $\delta_{CP}$           &  -1.789                             & [-2.450; -0.880]   \\  
 $\sin^{2}\theta_{13}$   &  0.0219                             & [0.0208; 0.0233]   \\  
 $\sin^2 \theta_{23}$    &  0.534                              & [0.490 ; 0.580]   \\  
 $\Delta m_{32}^2$       &   $2.539\times 10^{-3}$ eV$^2 \cdot$c$^{-4}$  & [-3.000; -2.952] $\times 10^{-3}$ eV$^2 \cdot$c$^{-4}$  \\
                         &                                     & [2.424; 2.664]$\times 10^{-3}$ eV$^2 \cdot$c$^{-4}$  \\  \hline\hline
\end{tabular}
\caption{Best-fit results and the 1$\sigma$ credible interval of the T2K data fit with the reactor constraint with the Bayesian analyses including both mass orderings.}
\label{tab:bestfit_worc_run1-7c_mach3}
\end{table}

In 2016-2017, T2K acquired $7.252 \cdot 10^{20}$ additionnal POT in neutrino mode, that are currently being analysed. 
The selection of the Super-K samples is being revised in order to increase the fiducial volume, that is excluding the volume 2 m from the inner detector wall in the analysis presented above.
Benefitting from a new reconstruction algorithm, this hard cut on the vertex location will be removed in order to observe an expected increase of 20\% more events, with a reduction of the background. 
The cross-section model is also being refined, notably by the addition of parameters controlling the uncertainties on short and long range correlations.
The T2K analyses are taking place in the prospect of the T2K-II phase, that plans to increase the total acquired POT to $20 \cdot 10^{21}$ in order to achieve a 3$\sigma$ exclusion of $\delta_{CP}$~\cite{t2k_2}.

\end{document}